# Handover between Macrocell and Femtocell for UMTS based Networks

Mostafa Zaman Chowdhury[†], Won Ryu[††], Eunjun Rhee[††], and Yeong Min Jang[†]
[†]Kookmin University, Korea
[††]Electronic and Telecommunications Research Institute (ETRI), Korea
mzceee@yahoo.com, {wlyu, ejrhee}@etri.re.kr, yjang@kookmin.ac.kr

*Abstract* — The femtocell networks that use home base station and existing xDSL or other cable line as backhaul connectivity can fulfill the upcoming demand of high data rate for wireless communication system as well as can extend the coverage area. Hence the modified handover procedure for existing networks is needed to support the macrocell/femtocell integrated network. Some modifications of existing network and protocol architecture for the integration of femtocell networks with the existing UMTS based macrocell networks are essential. These modifications change the signal flow for handover procedures due to different 2-tier cell (macrocell and femtocell) environment. The measurement of signal-to-interference ratio parameter should be considered for handover between macrocell and femtocell. A frequent and unnecessary handover is another problem for hierarchical network environment that must be optimized to improve the performance of macrocell/femtocell integrated network.

In this paper, firstly we propose the concentrator based and without concentrator based femtocell network architecture. Then we present the signal flow with appropriate parameters for the handover between 3GPP UMTS based macrocell and femtocell networks. A scheme for unnecessary handoff minimization is also presented in this paper. We simulate the proposed handover optimization scheme to validate the performance.

## 1. Introduction

The demand for high data rate for wireless communication is increasing tremendously. Existing wireless communication systems face many challenges to support high wide band data access. Both the coverage area and the capacity of existing cellular network systems are not sufficient to meet the expected demand of multimedia traffics. The closer of the transmitter and receiver of a wireless system cause increase capacity of a wireless link and creates dual benefits of higher quality links and more spatial reuse [13]. So among many approaches femtocell approach is one of the best approach to diverse the load from the cellular networks as well as to reduce the operating and capital expenditure costs for operators.

The network management and integration of femtocell with 3GPP macrocell networks is different from the existing 3GPP networks. Thousand of femtocells within a macrocell area create interference problem. Also due to huge number of possible target femtocell candidates for macrocell to femtocell handover need a large neighbor list and communication with many femtocells for the pre-handover procedure. The optimal solution of these two problems can improve the performance of femtocell networks. Hence the modifications of handover procedures for existing networks are needed.

A concentrator consists of femto gateway (FGW) and femtocell management system (FMS) in the femto access point (FAP) to core network connectivity can be used for the management of FAP. The FGW manages thousand of femtocells. Traffics from different femtocells come to FGW and then send to desired RNC and traffics come from RNC send to target femtocell. Also every FAP can have pre-registered user. Only this pre-registered user can access that FAP. Thus this technique can reduce the number of target femtocell for macrocell to femtocell handover. So communication signals among different FAP can be optimized.

The femtocell architecture is much more different than existing cellular networks. Thus the handover between macrocell and femtocell is one of the main issues for femtocell network deployment. A handsome amount of user all over the world use 3GPP based UMTS networks. Some modification of existing network architecture and protocol architecture for the integration of femtocell networks with the existing UMTS based macrocell networks is needed that also change the signal flow for handover procedures.

One of the serious problems for femtocell deployment is the incidence of unnecessary handover due to movement of the user. These unnecessary handovers cause the reduction of user's QoS level and system capacity. These unnecessary handovers can be optimized using proper call admission control (CAC) and resource management.

This paper is organized as follows. Section 2 provides the concentrator based femtocell network architecture. Call flow for handovers between macrocell and femtocell are presented in Section 3. In Section 4, a CAC has been proposed to minimize the unnecessary handover. The simulation results for the proposed handover minimization scheme are provided in section 5. Finally, we give our conclusion in Section 6.

## 2. Femtocell System Architecture

For the femtocell/macrocell network integration, there are several options are possible. Each option comes with a tradeoff in terms of scale but the best option depends on an operator's existing network capabilities and their future plan regarding the network expansion. Figure 1 shows the basic connectivity for femtocell network deployment [1], [10]-[11].

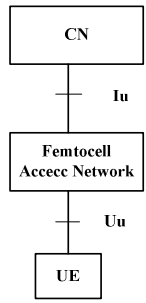

**Figure 1. Femtocell system architecture**

Traditional 2G/3G networks utilize centralized devices, RNCs, to control their associated base stations. One RNC in charge of radio resource management of about 100 base stations [12]. Within one macrocell coverage area there are thousand of femtocells. Thus a single RNC needs to control on the order of hundreds to thousands or tens of thousands of femtocells. It's not possible to handle or control so many FAPs using the current network control entities. Hence for femtocell deployment, FAP connectivity should be different than that of existing 3GPP network connectivity to improve the performance of overall system. Figure 2 is one of the candidates for device-to-core network connectivity for femtocell networks. Several FAPs are connected to FGW through broadband ISP network and $I_{u-h}$ interface. A new interface $I_{u-h}$ is introduced for femtocell network to connect the FAP over ISP network. The FGW acts like a concentrator. The concentrator node or FGW is connected to RNC over the standard $I_{ub}$ interface and looks like normal Node B.

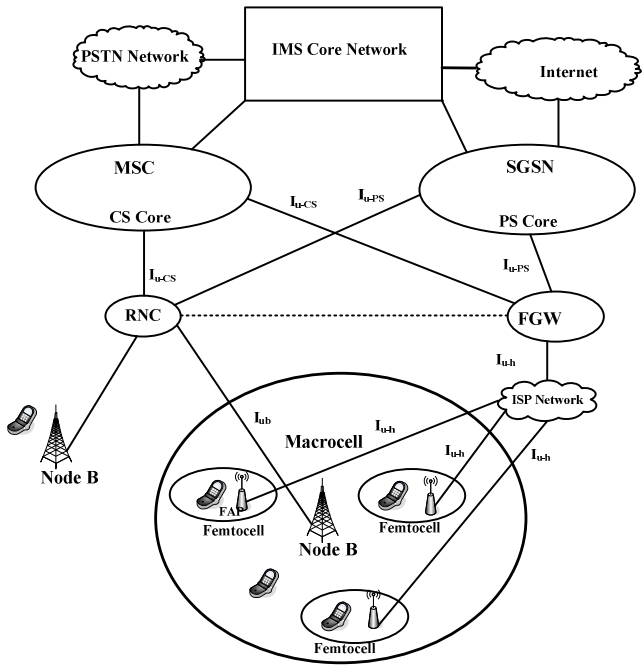

**Figure 2. Femtocell users to CN connectivity**

Figure 3 is another alternative architecture for femtocell deployment. This architecture is quite similar to the existing 3G-network architecture. Each femtocell in this architecture is considered as an equivalent of Node B and the FAP is connected to the RNC over an IP interface. The $I_{ub}$ signaling takes place over an IP network. Network security can be handled by the IP security protocol between the FAP and the security gateway node. This architecture is suitable for an operator who have an existing UMTS infrastructure deployed; the number of FAP within the macrocell is not much more; and who is looking for a fast integration of the femtocell into existing infrastructures. The problem of this architecture is the large number of FAP within a macrocell and broadcasting such large information through RNC incurs too much overhead.

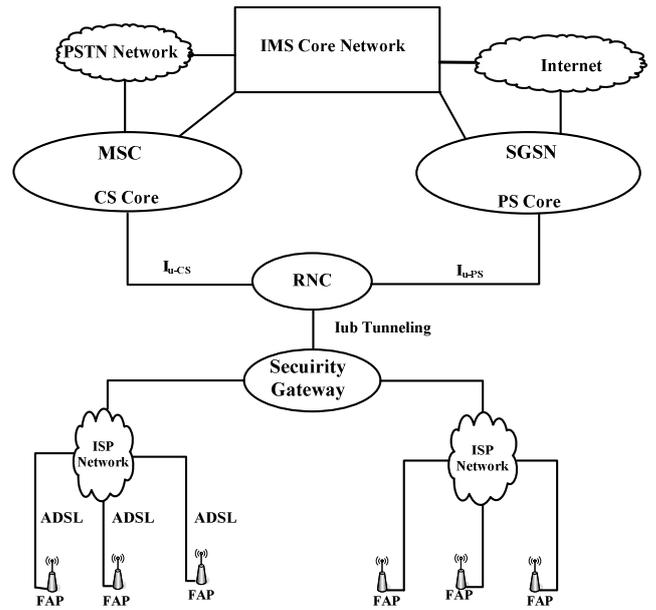

**Figure 3. Users to CN connectivity for fast integration of the femtocell into existing infrastructures**

## 3. Handover Call Flow between Macrocell and Femtocell

The ability to seamlessly switch between the femtocell and the macrocell networks is a key driver for femtocell network deployment. The handover procedures for existing 3GPP networks are presented in [2]-[8]. This section proposes the call flows based on the network architecture shown in Figure 2. The proposed handover schemes optimize the selection/reselection/RRC management functionalities in the femtocell/macrocell handover. The handover procedures are basically divided into two phases: handover preparation phase (information gathering, handover decision), and handover execution phase. During the information gathering phase, the MT collects information about the handover candidates, and authentications are acquired for security purposes. In handover decision phase, the best handover candidate is determined. Finally, after deciding to perform the actual handover, the mobile station (MS) initiates to connect with new AP. For the handover between macrocell and femtocell, initial network discovery for femtocell and initial access information gathering are needed. The most important difference between UMTS based macrocell networks and a femtocell network is the radio resource control (RRC) functionalities. FAP has the RRC functionalities whereas Node B has no RRC functionalities. So the proposed handover call flow for macrocell/femtocell integrated networks differs from that of existing 3GPP based macrocell networks.

## 3.1 Macrocell to Femtocell Handover

Macrocell to femtocell handover is the most challenging issue for femtocell network. Macrocell to femtocell handover is also complex compared to femtocell to macrocell handover. There are many possible target femtocells for handover. In this handover MS needs to select the appropriate target FAP among many candidate FAPs. Also interference level should the considered for handover decision. Serving Node B coordinates the handover of MS from Macro BS to a FAP by provisioning information of allowed FAPs (assuming that mostly closed FAPs are residing in) to scan for making a FAP neighbor list. Hence whenever the MS sends the measurement report to FAP, it should also inform about the interference level. The authorization should be checked during the handover preparation phase. Thus macrocell to femtocell handover consists more functionality than existing macrocell to macrocell handover. Figure 4 shows the call flow for the intra SGSN handover from UMTS based macrocell to femtocell network.

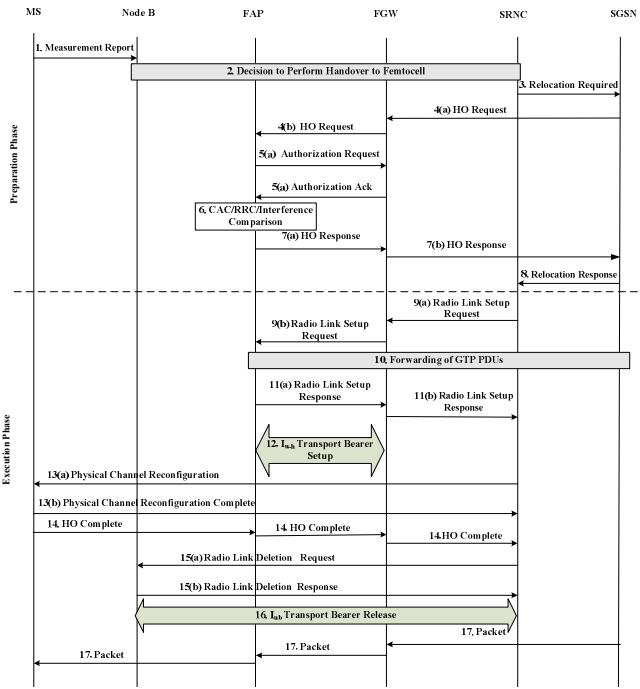

**Figure 4. Handover from macrocell to femtocell (Intra SGSN)**

## 3.2 Femtocell to Macrocell Handover

Figure 5 shows the call flow for the intra SGSN handover from femtocell to UMTS based macrocell network. The handover from femtocell to macrocell is not so complex like macrocell to femtocell handover because whenever a user move away from femtocell network, there is no option other than macrocell networks. The most important issue of this handover is that, the handover time should be very small. There is no complex interference calculation and authorization check in this handover like that of macrocell to femtocell handover.

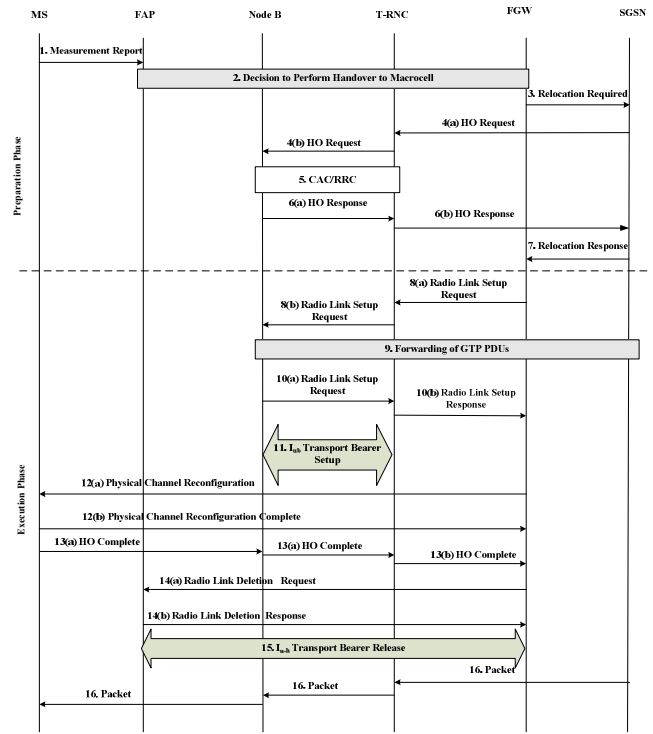

**Figure 5. Handover from femtocell to macrocell (Intra SGSN)**

## 4. Minimization of Unnecessary Handover

Frequent and unnecessary handover is another serious problem for femtocell networks environment as femtocell coverage area is very small and there is possibility to stay very small time whenever a high speed MS enters into femtocell coverage area. A high speed MS causes two unnecessary handover due to movement from macrocell area to femtocell area and again femtocell area to macrocell area. In the wireless communication systems, the frequent and unnecessary handovers reduce the end-to-end QoS level as well as decrease the capacity of the system. So the minimization of unnecessary handover is absolutely necessary for the femtocell/macrocell integrated network system in order to improve the user's QoS level and system capacity. Whenever a MS is connected with macrocell network and due to movement of MS, the MS found change of signal level from FAP. Sometimes MS with higher velocity cause very little time to stay in a femtocell coverage area. This causes unnecessary handovers that is indicated by "A" in Figure 6. In Figure 6 "B" indicates the case when a MS just move inside the femtocell coverage area and maintain good received signal level for long time. The "C" shown in Figure 6 indicates the case when a MS moves to femtocell area but does not enter into center area and stay at the boundary area for long time. Hence different types of condition arise. Due to arising of these different conditions, only a unique handover decision making policy is not sufficient to improve the performance. Thus some modified handover decision policy is needed.

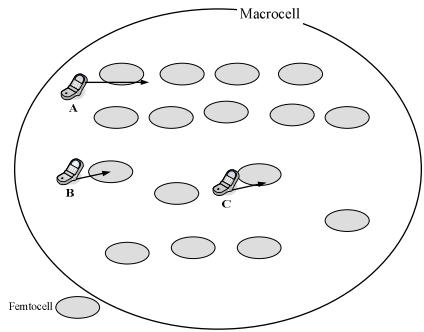

Figure 6. Movements of MS within macrocell/femtocell coverage area

### 4.1 Proposed CAC to Reduce Unnecessary Handover

Figure 7 shows the CAC to reduce the number of unnecessary handover whenever a macrocell user moves to femtocell coverage area.

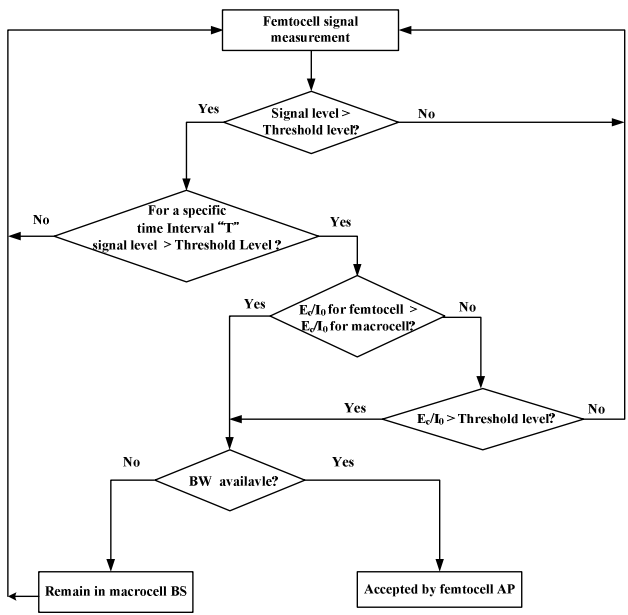

Figure 7. CAC to accept a handover call by FAP

Three parameters have been considered for this proposed CAC. These parameters are: received signal level, duration of time a MS maintains the minimum required signal level and signal-to-interference ($E_c/I_0$) level. The threshold level is the minimum level of signal that must be needed to handover a MS from macrocell to femtocell. The threshold time "T" will be specified by operator. Sometimes MS receives the signal greater than minimum required level but within very short time the level again go down. Whenever a MS moves from macrocell to femtocell, the MS must maintain minimum required threshold level of signal for minimum "T" time. Hence the threshold time will reduce the number of unnecessary handover. The interference is one of the most important issues for femtocell networks. Hence interference level is also considered for handover decision in proposed CAC.

## 5. Performance Evaluation

The performances of the proposed handover minimization scheme are performed using simulation result. Table 1 shows the basic simulation parameters. We calculate the angle of movement of a MS and then apparent stay time in the femtocell coverage area of that MS can be calculated form the velocity. We assume 100 femtocell APs within a macrocell coverage area. Then we make result for a single femtocell from the average result of all 100 femtocells.

Table 1. Simulation parameters

| Shape of femtocell coverage area | Circular |
|---|---|
| Radius of femtocell coverage area | 10 m |
| Average velocity of MS in femtocell coverage are | 0.9 km/hr |
| Average call life time after handoff from macrocell to femtocell | 90 sec |
| Call life time and user velocity | Exponential distribution |
| Number of FAP within a macrocell | 100 |

Figure 8 shows the number of handover from macrocell to femtocell and again femtocell to macrocell for different schemes. The larger value of the threshold time makes the lower number of handover. This scheme shows that, all the users move from macrocell to femtocell coverage area does not need to handover from macrocell to femtocell and again from femtocell to macrocell. Thus our proposed scheme optimized some unnecessary handoffs. Figure 8 also shows that a scheme without any proper CAC causes much more unnecessary handover than a CAC scheme with a threshold time of 20 seconds.

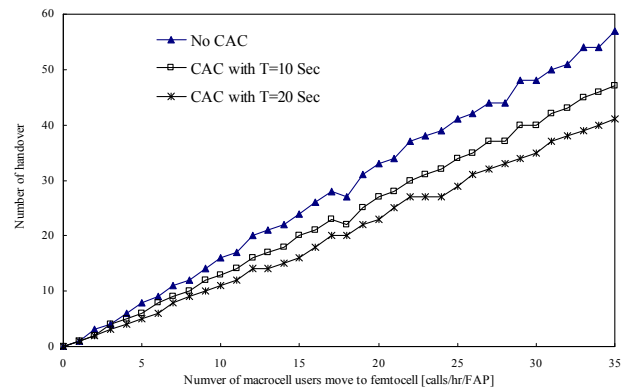

Figure 8. Observation of the number of handover for the users move from macrocell to femtocell coverage

Figure 9 shows the number of unnecessary handover minimization due to our proposed scheme. In our simulation, we consider a handover as an unnecessary handover when the MS move from macrocell to femtocell and within 60 seconds it moves to macrocell again or within 10 seconds it terminates the call. Figure 9 also shows that without any effective CAC, it makes about 38% unnecessary handover. A threshold time of 20 seconds and 10 seconds reduces the unnecessary handover into 8% and 19% respectively.

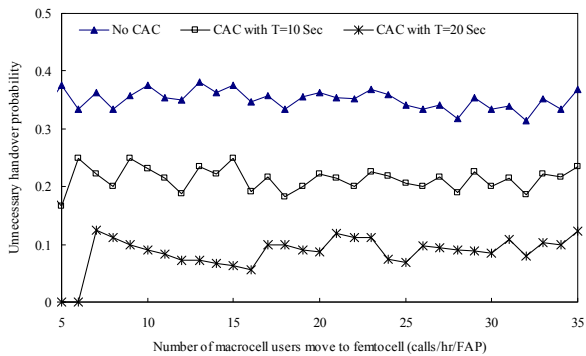

**Figure 9. Observation of unnecessary handover probability minimization for the users move from macrocell to femtocell**

## 6. Conclusion

Femtocell networks have the bright future to provide higher quality network access for indoor users at low price, while simultaneously reducing the burden on the whole network system. From a technical point of view, operators face challenges in providing a low cost solution, providing sufficient QoS over the IP backhaul, optimization of RF interference, mobility management and maintaining scalability. As femtocell coverage area is very small, there are some unnecessary handover occurs in macrocell to femtocell handover. The improvement of handover performances is depends on how the resources are handled for mobility management.

The proposed device to core network connectivity is able to make an efficient integration of femtocell networks with existing UMTS based macrocell networks. An efficient and reliable handover between macrocell and femtocell is possible using proposed handover schemes. The simulation results showed that the proposed unnecessary handover minimization scheme is an effective scheme to reduce the number of unnecessary handover. As femtocell is a very new but very promising technology and there is no complete specification yet, our proposed network architecture, call flow procedures and CAC might be very much helpful for future research direction about femtocell network deployment.

## Acknowledgement

This research was supported by the MKE (Ministry of Knowledge and Economy), Korea, under the ITRC (Information Technology Research Center) support program supervised by the IITA (Institute of Information Technology Assessment) (IITA-2008-C1090-0801-0019). This research was also supported by Electronic and Telecommunications Research Institute (ETRI), Korea.

## References

[1] 3GPP TR R3.020 V0.9.0, "Home (e) Node B: Network Aspects," September 2008.
[2] 3GPP TS 23.009 V5.11.0, "Handover Procedures," March 2006.
[3] 3GPP TS 23.060 V6.15.0, "General Packet Radio Service (GPRS) Service description," December 2006.
[4] 3GPP TR 25.931 V5.4.0, "UTRAN Functions, Examples on Signalling Procedures," June 2006.
[5] 3GPP TR 25.936 V4.0.1, "Handovers for Real-Time Services from PS Domain," December 2001.
[6] 3GPP TS 29.060 V6.18.0, "GPRS Tunnelling Protocol (GTP) across the Gn and Gp Interface," September 2007
[7] 3GPP TS 43.129 V7.2.0, "Packet-switched Handover for GERAN A/Gb Mode," May 2007.
[8] 3GPP TS 43.318 V8.1.0, "Generic Access Network (GAN)," February 2008.
[9] Frans Panken and Gerard Hoekstra, "Extending 3G/WiMAX Networks and Services through Residential Access Capacity," *IEEE Communications Magazine,* December 2007.
[10] http://www.femtoforum.org
[11] http://www.airwalkcom.com
[12] Shu-ping Yeh and Shilpa Talwar, "WiMAX Femtocells: A Perspective on Network Architecture, Capacity, and Coverage," *IEEE Communications Magazine*, October 2008.
[13] Vikram Chandrasekhar and Jeffrey G. Andrews, "Femtocell Networks: A Survey," *IEEE Communications Magazine,* September 2008.